\documentclass[twocolumn,12pt]{article}
\usepackage[T2A]{fontenc}
\usepackage{amsmath}
\usepackage{amssymb}
\usepackage{graphicx}
\usepackage{euscript}

\begin{document}
\title{Effect of dynamical traps on chaotic transport in a meandering jet flow}

\author{M.Yu. Uleysky, M.V. Budyansky, and S.V. Prants}
\maketitle
\begin{abstract}
We continue our study of chaotic mixing and transport of passive particles
in a simple model of a meandering jet flow [Prants, et al, Chaos
{\bf 16}, 033117 (2006)]. In the present paper we study and explain
phenomenologically a connection between dynamical, topological, and
statistical properties of chaotic mixing and transport in the model flow
in terms of dynamical traps, singular zones in the phase space where
particles may spend arbitrary long but finite time [Zaslavsky, Phys.~D
{\bf 168--169}, 292 (2002)]. The transport of passive particles is described
in terms of lengths and durations of zonal flights which are events
between two successive changes of sign of zonal velocity.
Some peculiarities
of the respective probability density functions for short
flights are proven to be caused by the so-called rotational-islands traps
connected with the boundaries of resonant islands (including 
the vortex cores) filled with the particles moving in the same frame and 
the saddle traps connected with  periodic saddle trajectories.
Whereas, the statistics of long flights can be
explained by the influence of the so-called ballistic-islands traps
filled with the particles moving from a frame to frame.
\end{abstract}

\section{Introduction}
{\bf Chaotic advection of water masses with their physical and biochemical
characteristics in quasi two-dimensional geophysical flows in the ocean and
atmosphere can be studied within the framework of Hamiltonian dynamics.
In a recent paper \cite{PBUZ06} we have
studied chaotic mixing and transport of passive particles in a simple kinematic model
of a meandering jet flow motivated by the problem of lateral
mixing in the western boundary currents in the ocean.
We found all the possible bifurcations of advection equations, described
the structure of the phase space (which is the physical space for
advected particles), and computed some statistical characteristics of
chaotic transport. In the present paper we establish a phenomenological
connection between dynamical, topological, and statistical properties of
chaotic transport and mixing in the same flow. Specific singular zones
in the phase space where particles
may spend arbitrary long but finite time (dynamical traps in
terminology by Zaslavsky \cite{Z02}), are responsible for anomalous statistical
properties. The dynamical traps, connected with rotational islands and saddle 
trajectories, are
responsible, mainly, for anomalous mixing, whereas those ones, connected with
ballistic islands~--- for anomalous transport. These dynamical traps may have
strong impact on transport and mixing in real geophysical jet flows.}

\centerline{\rule{0.4\textwidth}{1pt}}
Methods of the theory of dynamical systems are actively used to
describe advection of water (air) masses and their properties
in the ocean and atmosphere \cite{PBUZ06, S92, CNM93, NS97, YPJ02, 
KP06, D00, W05, Pi91, CLV99, SH03,UBP06,SMS89, BMS91, SWS93, SWS94,RW06}.
The geophysical jet currents, like the Gulf Stream, the Kuroshio,
the Antarctic circumpolar  current, and others in the ocean and
the polar night Antarctic jet in the atmosphere, are robust
structures whose form typically changes in space and time in a
meander-like way. If advected particles rapidly adjust their
own velocity to that of background flow and do not affect the flow
properties, then the particles are called passive (scalars, tracers,
or Lagrangian particles) and their equation of motion is very simple
\begin{equation}
\frac {d{\mathbf r}}{dt} = {\mathbf v}({\mathbf r},t)
\label{1}
\end{equation}
where ${\mathbf r}=(x, y, z)$ and ${\mathbf v}=(u, v, w)$ are the position vector
and the particle velocity at a point $(x, y, z)$. If the corresponding
Eulerian velocity field is supposed to be regular, the vector
Eq.~(\ref{1}) in nontrivial cases is a set of three nonlinear deterministic
differential equations whose phase space is a physical space of
advected particles. It is well known from dynamical systems
theory that solutions of this kind of equations can be chaotic in the
sense of exponential sensitivity to small variations in initial conditions
and/or control parameters.
As to advection equations, it was Arnold \cite{A65} who firstly suggested 
chaos in the field lines (and, therefore, in trajectories) for a special
class of three-dimensional stationary flows (so-called ABC flows), and this
suggestion has been confirmed numerically by H{\'e}non \cite{H66}. In the
approximation of incompressible planar flows, the velocity components
can be expressed in terms of a stream function \cite{Lamb}:
$u=-\partial\Psi/\partial y$ and $v=\partial\Psi/\partial x$. The equations of
motion (\ref{1}) in a two-dimensional incompressible
flow have now a Hamiltonian form with the streamfunction $\Psi$ playing the role
of a Hamiltonian and the coordinates $(x, y)$ of a particle playing the role
of canonically conjugated variables.

Advection of passive particles has been shown to be chaotic in a number
of theoretical \cite{W05, Pi91, CLV99,SH03,PBUZ06, S92, CNM93, NS97, YPJ02,UBP06,KP06,
RW06} and laboratory
\cite{SMS89, BMS91, SWS93, SWS94} models of geophysical jet currents.
In our recent paper \cite {PBUZ06}~(Paper I) we have studied mixing,
transport, and chaotic advection in a simple kinematic model of a meandering
two-dimensional jet flow with a Bickley zonal velocity profile
$u\sim\operatorname{sech}^2 y$ being motivated by the problem of lateral
mixing of water masses (together with salinity, heat, nutrients,
pollutants, and other passive scalars) in the western boundary currents
in the ocean. We derived advection equations in the frame moving with
the phase velocity of a running wave imposed on the Bickley jet,
found the stationary points, conditions of their stability, and all the
possible bifurcations of these equations which were shown to be autonomous
in the co-moving frame. Under a periodic perturbation of the wave
amplitude, the phase plane of the chosen model flow has been
shown to consist of a central eastward jet, peripheral westward currents,
and chains of northern and southern circulations (vortex cores)
immersed in a chaotic sea which, in turn, contains islands of regular
motion. Statistical properties of chaotic transport of advected particles
have been characterized in terms of particle's zonal flights (any
event between two successive changes of the sign of the particle's
zonal velocity $u$). Probability density functions (PDFs) of
durations and lengths of flights, computed with a number of very
long chaotic trajectories, have been found to be complicated
functions with local maxima and fragments with exponential and
power-law decays.

The aim of this paper is to study a phenomenological connection
between dynamical, topological, and statistical properties of chaotic
mixing and transport in the meandering jet flow considered in Paper~I
and to explain transport properties by phenomenon of the so-called
dynamical traps. Following to Zaslavsky \cite{Z02}, the dynamical trap
is a domain in the phase space of a Hamiltonian system where a particle
(or, its trajectory) can spend arbitrary long finite time, performing
almost regular motion, despite the fact that the full trajectory is chaotic
in any appropriate sense.  In fact, it is the definition of a quasi-trap.
Absolute traps, where particles could spend an infinite time, are
possible in Hamiltonian systems only with a zero measure set. 
The dynamical traps are due to a
stickiness of trajectories to some singular domains in the phase space,
largely, to the boundaries of resonant islands, saddle trajectories, and cantori.
There are no classification and description of the dynamical traps.
Zaslavsky described two types of dynamical traps in Hamiltonian systems:
hierarchical-islands traps around chains of resonant islands
\cite{Z95, ZN97, Z05} and stochastic-layer traps which are stochastic jets
inside a stochastic sea where trajectories can spend a very long time
\cite{P90,Z02,Z05}.
It is expected that classification and description of the most
typical dynamical traps would help us to construct kinetic equations which
will be able to describe transport properties of chaotic systems including
anomalous ones \cite{Z02, Za02,Z05}.

This paper is organized as follows. We start with advection equations derived
in Paper~I in the frame of reference moving with the phase velocity
of a meander whose amplitude changes in time periodically. We compute
in Sec.~II PDFs for lengths $x_f$ and durations $T_f$ of zonal
flights for a number of chaotic trajectories and show analytically that
all the flights start and finish only inside a strip confined
by two curves whose form is defined by the condition $u=0$.
Some prominent peaks in statistics of short flights ($|x_f|<2\pi$)
are proved to be caused by stickiness of trajectories to the
boundaries of rotational resonant islands filled with
regular particles rotating in the same frame (Sec.~III).
We call this type of dynamical traps as a rotational-island trap (RIT).
In Sec.~IV we study dynamical traps connected with periodic saddle 
trajectories, emerged from saddle points of the unperturbed system (\ref{4}) 
under the perturbation (\ref{5}), and prove numerically that the saddle traps 
(STs) contribute to the statistics of short flights as well. Another type of islands, ballistic islands (filled with regular
particles moving from frame to frame), is proved to contribute
to the statistics of long flights ($|x_f|\gg 2\pi$) in Sec.~V. Both the
ballistic-islands trap (BITs) and RITs belong to the class
of hierarchical-island traps by the Zaslavsky's classification.

\section{\label{Description}Description of chaotic transport in terms of flights}
\subsection{Basic features of the flow}
We take the following specific stream function as a kinematic
model of a meandering jet flow in the laboratory frame of reference:
\begin{multline}
\Psi'(x',y',t')=\\
=-\Psi'_0\tanh{\left(\frac{y'-a\cos{k(x'-c t')}}
{\lambda\sqrt{1+k^2 a^2\sin^2{k(x'-c t')}} } \right)},
\label{2}
\end{multline}
where the width of the jet is $\lambda$. Meandering is provided
by a running wave with the amplitude $a$, the wave-number $k$, and the
phase velocity $c$. The normalized streamfunction in the frame moving
with the phase velocity is
\begin{equation}
\Psi=-\tanh{\left(\frac{y-A\cos x}{L\sqrt{1+A^2\sin^2 x}}\right)}+Cy,
\label{3}
\end{equation}
where $x=k(x'-ct')$ and $y=ky'$ are new scaled coordinates, and
$A=ak$, $L=\lambda k$, and $C=c/\Psi'_0 k$ are the control parameters.
Equations, governing advection of passive particles (\ref{1}) in the
co-moving frame, are the following:
\begin{equation}
\begin{gathered}
\begin{aligned}
\dot x&=\frac{1}{L\sqrt{1+A^2\sin^2 x}\cosh^2\theta}-C,\\
\dot y&=-\frac{A\sin x(1+A^2-Ay\cos x)}{L\left(1+A^2\sin^2 x\right)^{3/2}
\cosh^2\theta},
\end{aligned}\\
\theta=\frac{y-A\cos x}{L\sqrt{1+A^2\sin^2 x}},
\end{gathered}
\label{4}
\end{equation}
where dot denotes differentiation with respect to dimensionless time
$t=\Psi'_0 k^2 t'$.

\begin{figure}[!htb]
\centerline{
\includegraphics[width=0.48\textwidth,clip]{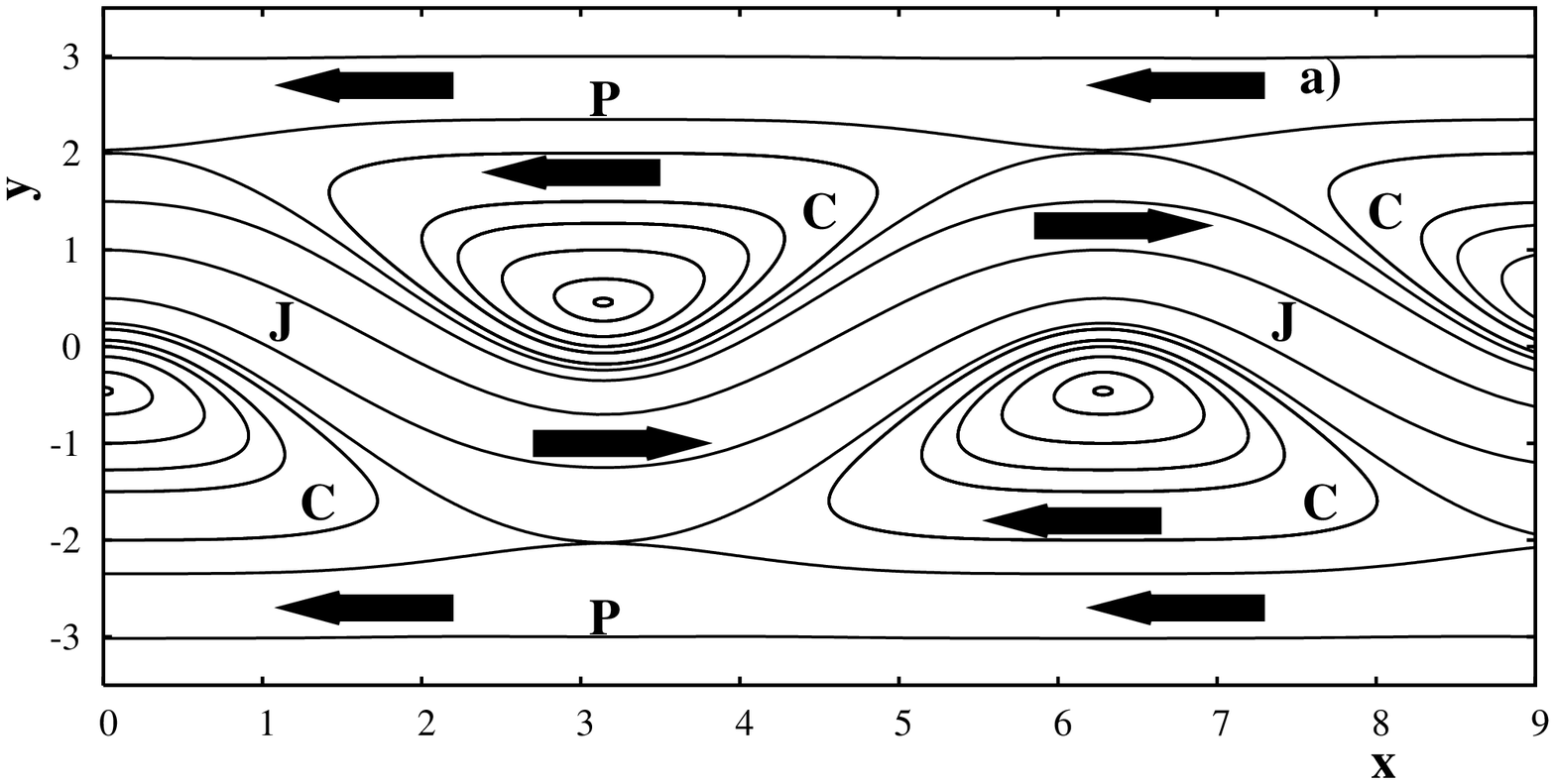}}
\centerline{
\includegraphics[width=0.48\textwidth,clip]{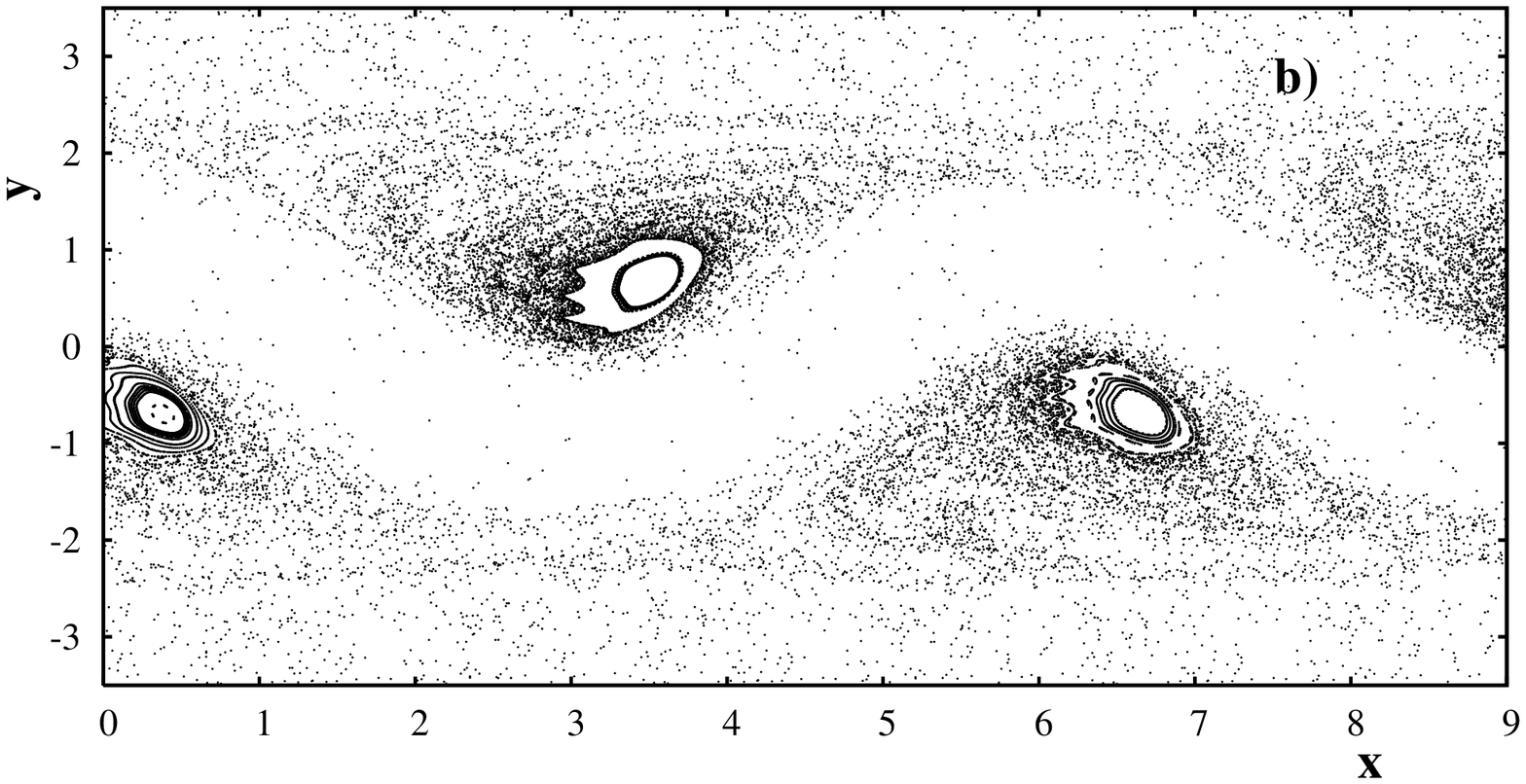}}
\centerline{
\includegraphics[width=0.48\textwidth,clip]{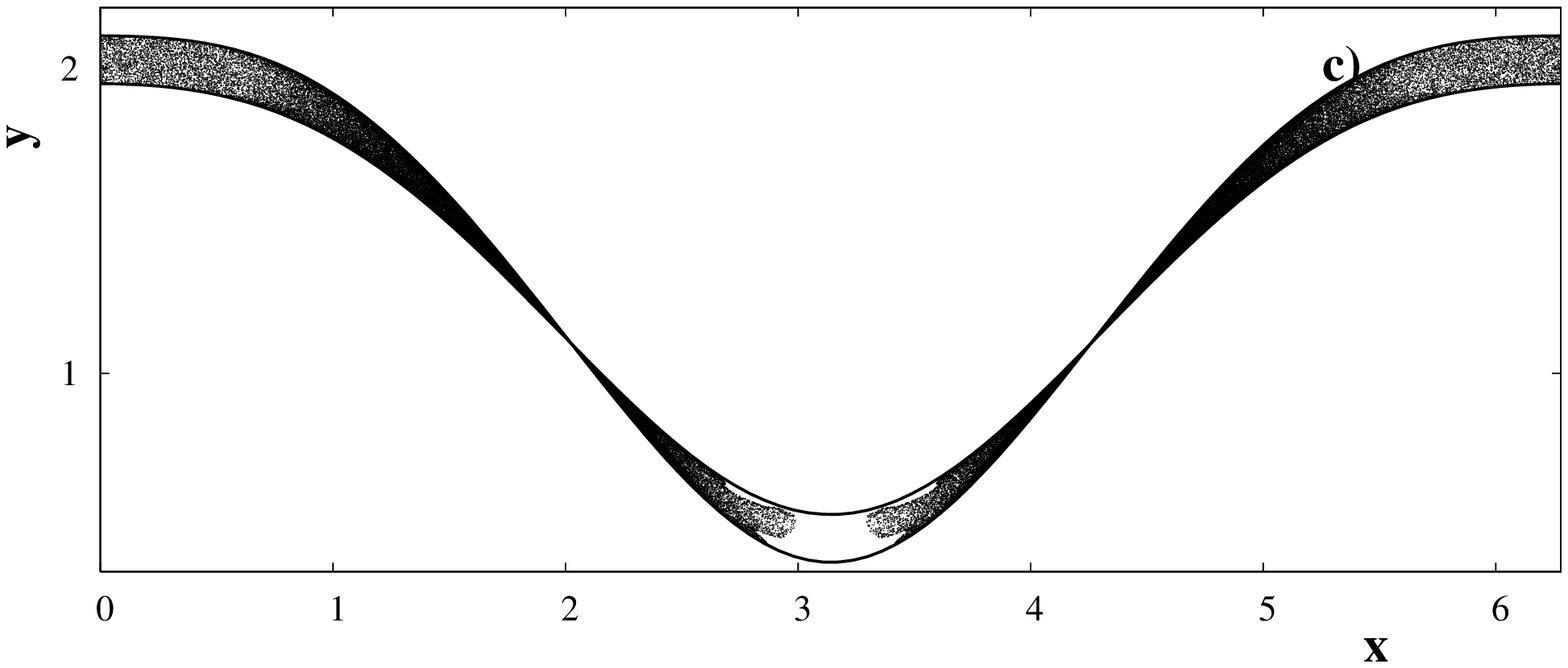}}
\caption{(a) Stationary streamfunction of a meandering jet in the
co-moving frame (\ref{3}). The flow is divided into three
different regimes: circulations ($C$), jet ($J$), and peripheral currents ($P$).
(b) Poincar{\'e} section of the perturbed meandering jet in the co-moving
frame. The parameters of the steady flow are: the jet's width
$L=0.628$, the meander's amplitude $A_0=0.785$ and its phase velocity
$C=0.1168$. The perturbation amplitude and frequency are:
$\varepsilon=0.0785$ and $\omega=0.2536.$
(c) Turning points of a single
chaotic trajectory on the cylinder $0\le x\le 2\pi$ are in a strip confined
by two curves (\ref{6}) with $A=A_0\pm \varepsilon$.}
\label{fig1}
\end{figure}

In Paper~I (for more details see \cite{UBP06}) we have found and
analyzed all the stationary points, their stability, and the
bifurcations of the equations of motion (\ref{4}).
Being motivated by the problem of mixing and transport of water masses
and their properties in oceanic western boundary currents like the
Gulf Stream and the Kuroshio, we chose the phase portrait shown in
Fig.~\ref{fig1}a among all the possible flow regimes.
Passive particles can move along stationary (in the co-moving frame)
streamlines in a different manner. They can move to the east in the jet ($J$)
and to the west in northern and southern (with respect to the jet)
peripheral currents ($P$). There are also particles rotating in the northern
and southern circulation cells (C) in a periodic way. 
The northern separatrix connects the saddle points
at $x_s^{(n)}=2\pi n$ and $y_s^{(n)}=L\operatorname{Arcosh}
\sqrt{1/LC}+A$
and the southern one connects the saddle points at $x_s^{(s)}=(2n+1)\pi$
and $y_s^{(s)}=-L\operatorname{Arcosh}\sqrt{1/LC}-A$, where
$n=0, \pm1, \dots$.

As a perturbation, we took in Paper~I the simple periodic modulation of the
meander's amplitude
\begin{equation}
A=A_0+\varepsilon\cos(\omega t + \phi).
\label{5}
\end{equation}
Under the perturbation, there arise resonances between the perturbation
frequency $\omega$ and the frequencies $f$ of the particle's rotation in the
circulations $C$. A frequency map $f(x_0,y_0)$, computed in Paper~I
(see Fig.~2 in that paper), shows the values of $f$ for particles with
initial positions $(x_0, y_0)$ in the unperturbed flow. With a given value
of the perturbation frequency and fixed values of the other control parameters,
the vortex cores in the circulations survive,  stochastic layers
appear along the unperturbed separatrix, and the central jet $J$ is a
barrier to transport of particles across the jet.
In Paper~I we fixed the scaled values of the parameters of the unperturbed
flow, the jet's width $L=0.628$, the meander's amplitude $A_0=0.785$
and its phase velocity $C=0.1168$, that are in the range of the
realistic values for the Gulf Stream \cite{B89, BR89}, and took the initial 
phase to be $\phi=\pi/2$. The perturbation
frequency $\omega=0.2536$ chosen in Paper~I is close to the values
of the rotation frequency $f$ of the particles circulating in the inner
core of the regions $C$ (see Fig.~2 in Paper~I). In
Fig.~\ref{fig1}~b we show the Poincar{\'e} section (for a large number
of trajectories) of the meandering jet whose amplitude is modulated
with the frequency $w=0.2536$ and the strength $\varepsilon=0.0785$.
The vortex cores survive under this perturbation, the stochastic layers
appear along the unperturbed separatrix, and a central jet $J$ is a
barrier to transport of particles across the jet.

The equations of motion (\ref{4}) with the perturbation (\ref{5}) are
symmetric under the following transformations: (1) $t\to t$,
$x\to\pi+x$, $y\to -y$ and (2) $t\to -t$, $x\to\ -x$, $y\to y$.
It implies that the meridional transport (north-south and south-north)
is symmetric but the zonal transport (west-east and east-west) is
symmetric under a time reversal. Due to these symmetries motion can be
considered on the cylinder with $0\le x\le 2\pi$ and $y\ge 0$.
The part of the phase space with $2\pi n\le x\le 2\pi (n+1)$,
$n=0, \pm 1, \dots$, is called a frame.

It is convenient to characterize chaotic mixing and transport in terms
of zonal flights. A zonal flight is a motion of a particle between two
successive changes of signs of its zonal  velocity, i.~e. the motion between two
successive events $\dot x=u=0$.
Particles (and corresponding trajectories) in chaotic jet flows can be
classified in terms of the lengths of flights $x_f$ as follows.
The trajectories with $|x_f|<2\pi$ correspond to the particles moving
in the same frame or in neighbor frames. In the global stochastic layer 
there are particles moving
chaotically forever in the same frame but they are of a zero measure.
Among the particles with inter-frame motion, there are regular and chaotic
ballistic ones. Regular ballistic trajectories can be defined as those
which cannot have two flights with $|x_f|>2\pi$ in succession.
They correspond to particles moving in regular regions of the phase space
persisting under the perturbation, (eastward motion in the jet and
western motion in the peripheral current) and those moving in the
stochastic layer (trajectories belonging to ballistic islands).
Typical chaotic trajectories have complicated
distributions over the lengths and durations of flights.

In the laboratory frame of reference, all the fluid particles move to 
the east together 
with the jet flow and a flight is a motion between two 
successive events when the particle's zonal velocity $U$ is equal to the 
meander's phase velocity $c$. If $U<c$, the corresponding particle is 
left behind the meander (it is a western flight in the co-moving frame), 
if $U>c$, it passes the meander (an eastern flight in the co-moving frame). 
Short flights with $|x_f| < 2\pi$ (motion in the same spatial frame in the 
co-moving frame of reference) correspond to 
the motion in the laboratory frame when two successive events 
$U=c$ occur on the space interval 
less than the meander's spatial period $2\pi/k$. Ballistic flights between 
the spatial frames in the co-moving frame with $|x_f| > 2\pi$ correspond to 
the motion in the laboratory frame when the  particles move 
through more than one meander's crest between two successive events $U=c$.

\subsection{Turning points}
As in Paper~I, we will characterize statistical properties of
chaotic transport  by probability density functions (PDFs) of lengths of
flights $P(x_f)$ and durations of flights $P(T_f)$ for a number of
very long chaotic trajectory.
Both regular and chaotic particles may change many times the sign
of their zonal velocity $\dot x=u$.
From the condition $\dot x=0$ in Eq.~(\ref{4}), it is easy to find the
equations for the curves which are loci of turning points
\begin{multline}
Y_{\pm}(x,A)=\pm L\sqrt{1+A^2\sin^2{x}}\times\\
\times \operatorname{Arsech}{\sqrt{{LC\sqrt{1+A^2\sin^2{x}}}}}+A\cos x.
\label{6}
\end{multline}
We consider the northern curve, i.~e., Eq.~(\ref{6}) with the positive sign.
Taking into account that the perturbation has the form (\ref{5}), we
realize that all the northern turning points are inside a strip confined
by two curves of the form (\ref{6}) with $A=A_0\pm \varepsilon$.
Let us analyze the derivative over the varying parameter A
\begin{multline}
\frac{\partial Y}{\partial A}=\cos x+\\
+\frac{ACL^2 \sin^2 x}{2D}\biggl(2\operatorname{Arsech}{\sqrt{D}}-\frac{1}{\sqrt{1-D}}\biggr),
\label{7}
\end{multline}
where $D=LC\sqrt{1+A^2\sin^2 x}$. If the derivative at a fixed value of $x$
does not change its sign on the interval
$A_0-\varepsilon\le A\le A_0+\varepsilon$, then $Y$ varies from
$Y(x, A_0-\varepsilon)$ to $Y(x, A_0+\varepsilon)$, and for
each value of $y$ we have a single value of the perturbation parameter $A$.
However, there may exist such values of $x$ for which the equation
$\partial Y/\partial A=0$ has a solution on the interval
mentioned above. In this case one may
have more than one values of $A$ for a single value of $y$. Thus, the width
of the strip, containing turning points, is defined by the values of
$Y$ at the extremum points and at the end points of the interval of the
values of $A$. In Fig.~\ref{fig1}~c we show the turning points of a single
chaotic trajectory on the cylinder $0\le x\le 2\pi$ confined between two
corresponding curves.

\begin{figure}[!htb]
\centerline{\includegraphics[width=0.48\textwidth,clip]{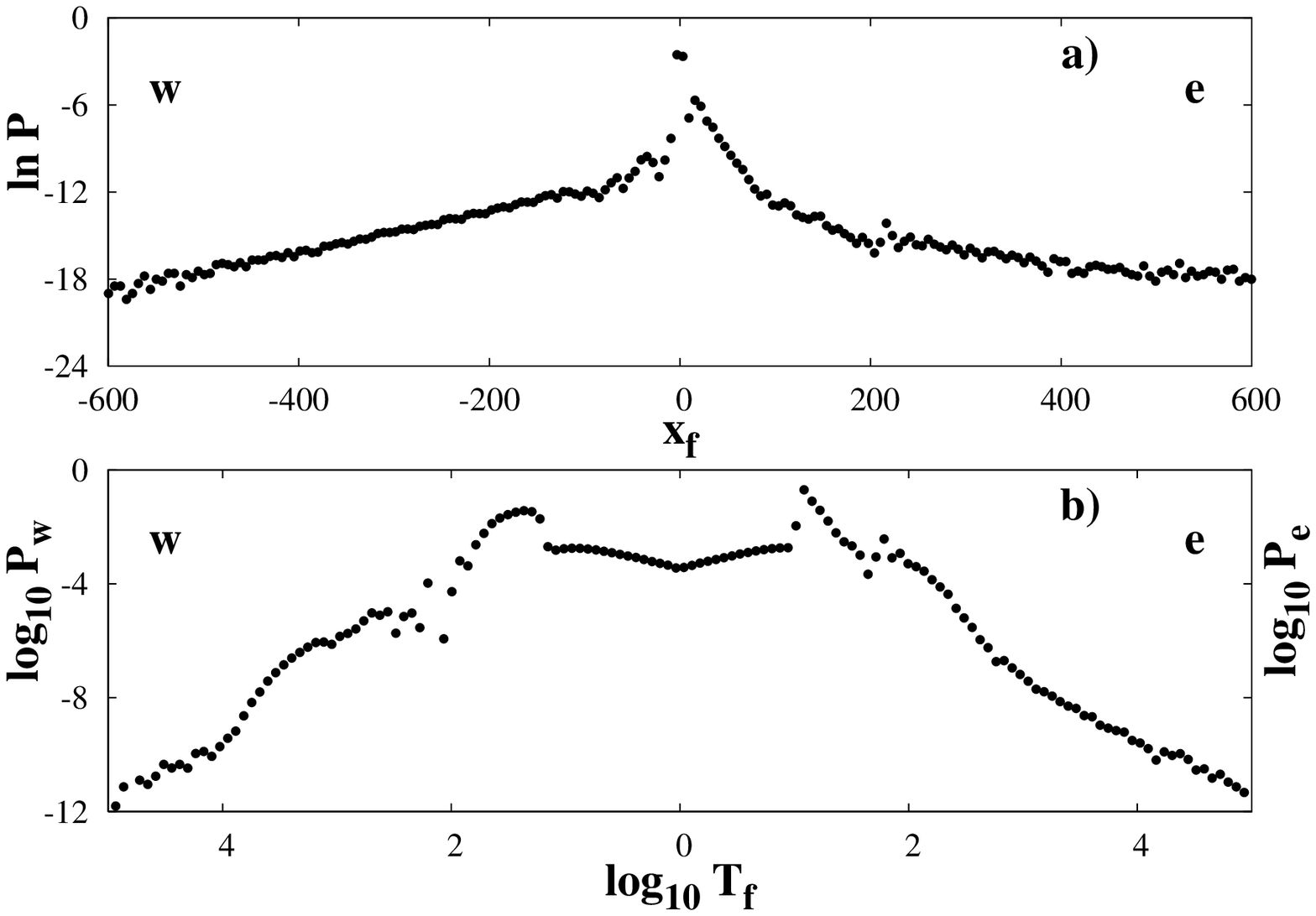}}
\caption{Probability density functions of (a) lengths $x_f$ and (b) 
durations $T_f$ of the westward (w) and eastward
(e) flights. The PDFs $P_{\rm w}(T_f)$ and $P_{\rm e}(T_f)$ are normalized
to the number of westward ($4.23 \cdot 10^7$) and eastward 
($4 \cdot 10^7$) flights, respectively. Statistics for five tracers 
with the computation time $t=5\cdot 10^8$ for each one.}
\label{fig2}
\end{figure}

In the numerical simulation throughout the paper we use the Runge-Kutta  
integration scheme of the fourth order with the constant time 
step $\Delta t 0.0247$.  
To study chaotic transport  we have carried out numerical experiments with tracers initially
placed in the stochastic layer. It was found that statistical properties
of chaotic transport practically do not depend on the number of tracers
provided that the corresponding trajectories are sufficiently
long ($t\simeq 10^8$). The PDFs for the lengths $x_f$ and durations $T_f$ of
flights for five tracers with the computation time $t=5\cdot 10^8$ for
each tracer are shown in Figs.~\ref{fig2}~a and b, respectively, both
for the eastward (e) and westward (w) motion.
Both $P(x_f)$ and $P(T_f)$ are complicated functions with local extrema
decaying in a different manner for different ranges of $x_f$ and $T_f$. The main
aim of our study of chaotic transport is to figure out the basic peculiarities
of the statistics and attribute them to specific zones in the phase space,
namely, to dynamical traps strongly influencing the transport.

\section{Rotational-islands traps}

It is well known, that in nonlinear Hamiltonian systems a complicated structure
of the phase space with islands,
stochastic layers, and chains of islands, immersed in a stochastic sea,
arises under a perturbation due to a variety of nonlinear resonances and
their overlapping \cite{C79}. The motion is quasiperiodic and stable in the
islands. The boundaries of the islands are absolute barriers to transport:
particles can not go through them neither from inside nor from
outside. 
Invariant curves of the unperturbed system (see Fig.~\ref{fig1}~a) are
destroyed under the perturbation (\ref{5}) (see Fig.~\ref{fig1}~b). 
As the perturbation strength
$\varepsilon$ increases, a closed invariant curve with frequency
$f$ is destroyed at some critical value of $\varepsilon$.
If the $f/\omega$ is a rational number, the corresponding curve
is replaces by an island chain, while the curves with
irrational frequencies are replaced by cantori 
(for a review see \cite{Mackay}). There are
uncountably many cantori forming a complicated hierarchy.
Numerical experiments with a variety of Hamiltonian systems
with different number of degrees of freedom provide an evidence for
the presence of strong partial barriers to transport
around the island's boundaries (for review, see \cite{Z05})
which manifest themselves on Poincar{\'e} sections as domains
with increased density of points.

In Paper~I we have found that with chosen values of the control parameters
there exist in each frame a vortex core (which is an island of the
primary resonance $\omega=f$) immersed into a stochastic sea, where there
are six islands of a secondary resonance emerged from three islands
of the primary resonance $3f=2\omega$ (see Fig.~3 in Paper~I).
Chains of smaller size islands are present around the vortex core and
the secondary-resonance islands. Particles belonging
to all of these islands (including the vortex core) rotate in the same frame
performing short  flights with the lengths $|x_f|<2\pi$. So we will
call them {\it rotational islands} and distinguish from the so-called {\it ballistic
islands} to be considered below.

\begin{figure}[!htb]
\centerline{\includegraphics[width=0.482\textwidth,clip]{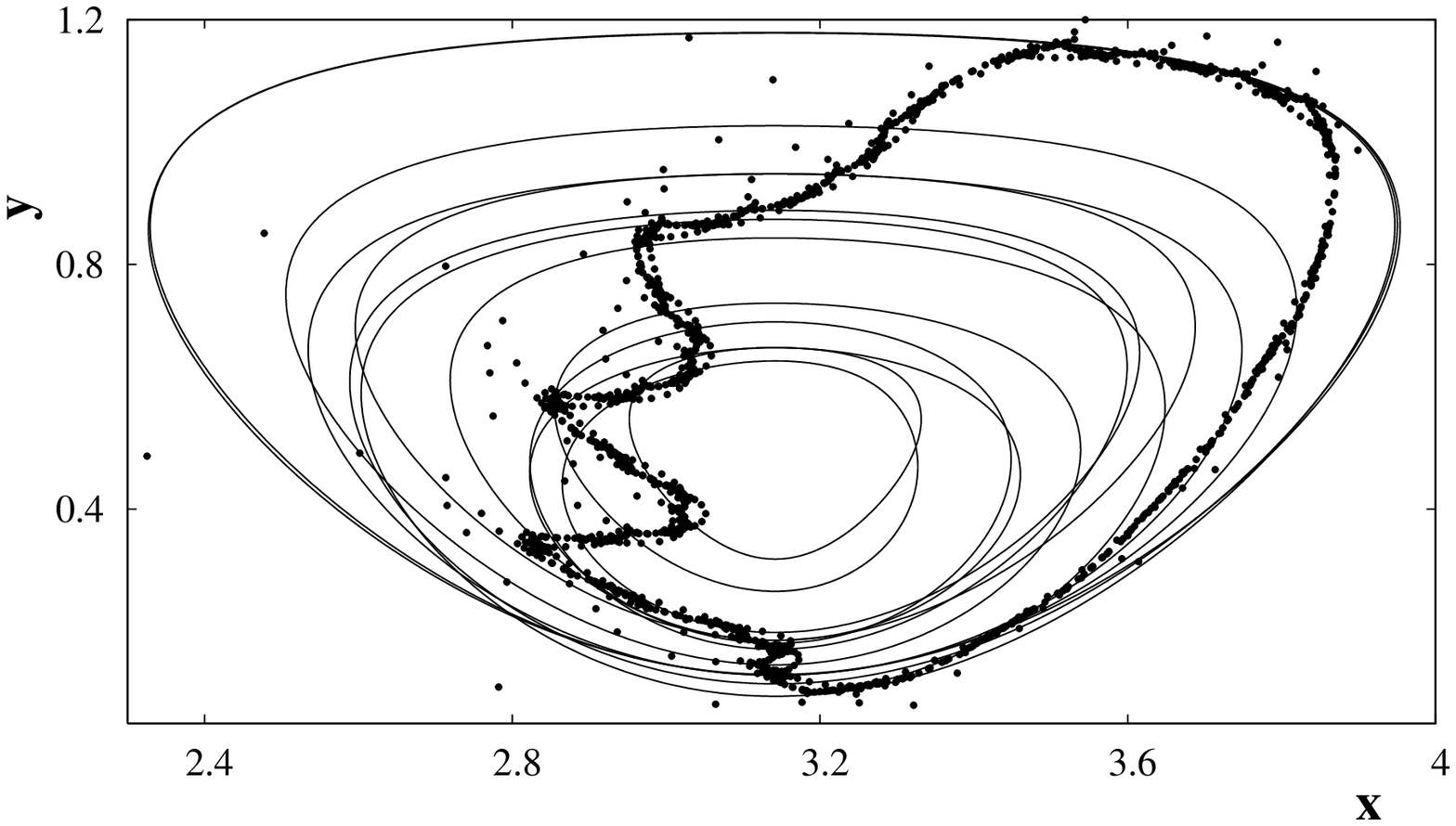}}
\caption{The vortex-core trap. Poincar{\'e} section of a
chaotic trajectory in the frame $0\le x\le 2\pi$ with a
fragment of a trajectory.}
\label{fig3}
\end{figure}

\begin{figure}[!htb]
\centerline{\includegraphics[width=0.482\textwidth,clip]{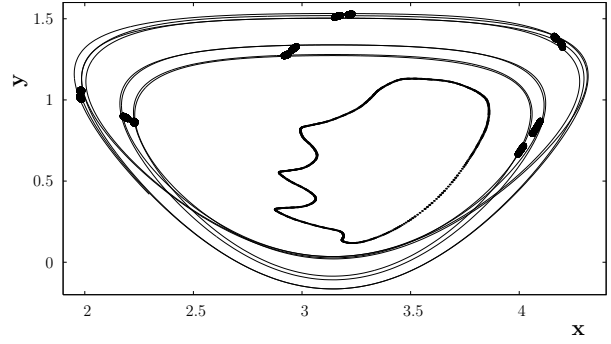}}
\caption{The secondary resonance islands trap. A fragment of a
chaotic trajectory sticking to the islands is shown.}
\label{fig4}
\end{figure}

Stickiness of particles to boundaries of the rotational islands has been
demonstrated in Paper~I. It means that real fluid particles can be trapped
for a long time in a singular zone nearby the borders of the rotational
islands which we will call
{\it rotational-islands traps} (RITs). To illustrate the effect of the RITs we
demonstrate in Figs.~\ref{fig3} and \ref{fig4} the Poincar{\'e}
sections of a chaotic trajectory in the frame $0\le x\le 2\pi$
sticking to the vortex core and to the secondary-resonance islands,
respectively. The contour of the vortex core is shown in Fig.~\ref{fig4} 
by the thick line. The small points are tracks of the particle's
position at the moments of time $t_n=2\pi n/\omega$ (where $n=1,2,\dots$)
and the thin curves are fragments of the corresponding trajectory
on the phase plane. Increased density of points indicates the presence of
dynamical traps near the boundaries of the rotational islands. Contribution of
the vortex-core RIT (Fig.~\ref{fig3}) to chaotic transport is expected to be much more
significant than the one of the RITs of the other islands
(Fig.~\ref{fig4}).
\begin{figure}[!htb]
\centerline{\includegraphics[width=0.48\textwidth,clip]{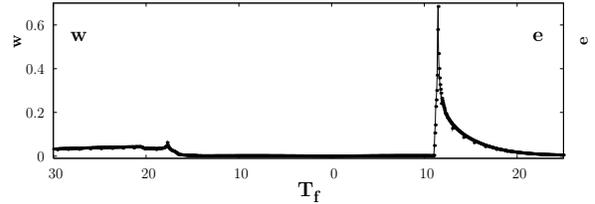}}
\caption{The PDFs for the eastward (e) and westward (w)
flights with the length shorter than $2\pi$. 
The PDFs $P_{\rm w}(T_f)$ and $P_{\rm e}(T_f)$ are normalized
to the number of westward ($4.19 \cdot 10^7$) and eastward 
($3.7 \cdot 10^7$) flights, respectively. Statistics for five tracers 
with the computation time $t=5\cdot 10^8$ for each one.} 
\label{fig5}
\end{figure}
\begin{figure}[!htb]
\centerline{\includegraphics[width=0.48\textwidth,clip]{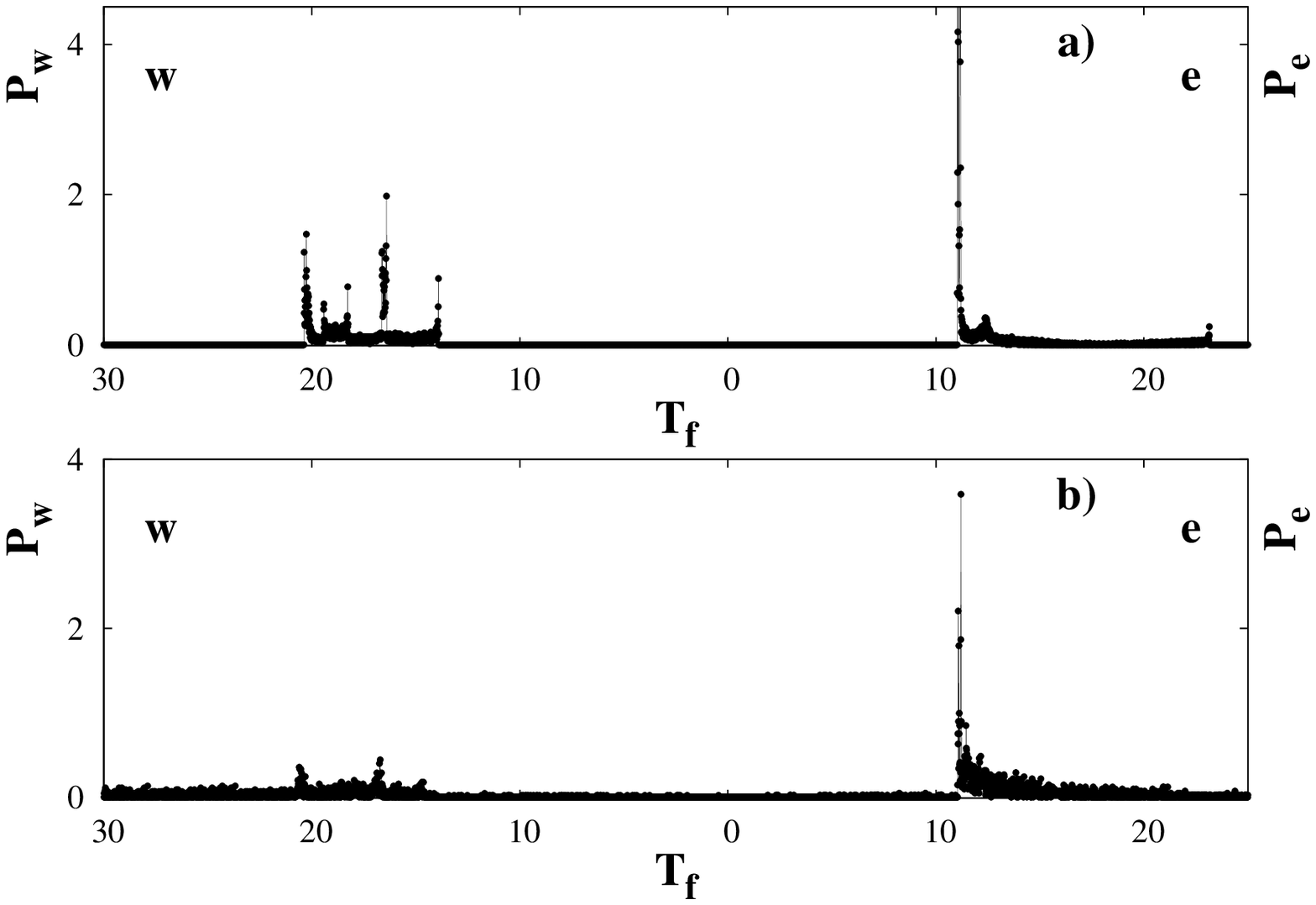}}
\caption{The vortex-core trap PDFs of durations $T_f$ of the
eastward (e) and westward (w) flights. 
(a) Regular quasiperiodic trajectory with the duration $t=2\cdot 10^5$  
inside the vortex core close to 
its boundary. Both the PDFs are normalized
to the number $8 \cdot 10^3$ of corresponding flights.  (b) Chaotic 
trajectory with the duration $t=2\cdot 10^5$ sticking to the boundary of the vortex core
from the outside. Both the PDFs are normalized
to the number $4 \cdot 10^3$ of corresponding flights.}
\label{fig6}
\end{figure}

It is reasonable to suppose that RITs contribute to the statistics of
short flights. By short flights we mean the flights with the length shorter
than $2\pi$. In Fig.~\ref{fig5} we show the part of the full PDF $P(T_f)$
(Fig.~\ref{fig2}~b) for the eastward (e) and westward (w) short flights
separately. There are a comparatively small number of the eastward flights
with $T_f < 11$. Let us note 
the prominent peak of the corresponding PDF at $T_f\simeq 11$ 
followed by an exponential decay. As to the westward short flights, there are
two small local peaks around $T_f\simeq 17$ and $21$.

To estimate the contribution of the vortex-core RIT to the statistics of short
flights, we compute and compare the statistics of the durations of flights
$T_f$ for two trajectories: a regular quasiperiodic one with the
initial position close to the inner border of the vortex core
(Fig.~\ref{fig6}~a) and a chaotic one with the initial position close to the
vortex-core border from the outside (Fig.~\ref{fig6}~b). Each full
rotation of a particle in a frame consists of two flights, eastward and
westward, with different values of $T_f$ because of the zonal asymmetry of
the flow. The statistics for the chaotic trajectory, sticking to the vortex core
(Fig.~\ref{fig6}~b), may be considered as a distribution of the durations of
flights in the vortex-core RIT.  The minimal flight duration in this RIT is
$T_f\simeq 11$
(the flights with smaller values of $T_f$ are rare and they occur outside the
trap). Positions of the local maxima of the PDF for the sticking trajectory
in Fig.~\ref{fig6}~b correlate approximately with the corresponding local maxima
of the PDF for the regular trajectory inside the core in Fig.~\ref{fig6}~a.
The similar correlations have been found (but not shown here) between
the local maxima of the PDFs for the lengths of flights $P(x_f)$ for the
interior regular and sticking chaotic trajectories.
These correlations and positions of the peaks prove numerically that short
flights with $|x_f|<2\pi$ and $11\lesssim T_f\lesssim 21$
may be caused by the effect of vortex-core RIT. We conclude from
Fig.~\ref{fig6}~b that the vortex-core RIT contributes to the statistics of
the short flights in the range $11\lesssim T_f\lesssim 20$
for the eastward flights with the prominent peak at $T_f\simeq 11$ and in
the range $15\lesssim T_f\lesssim 21$
for the westward flights with small peaks
at $T_f\simeq 17$ and $21$.

\begin{figure}[!htb]
\centerline{\includegraphics[width=0.48\textwidth,clip]{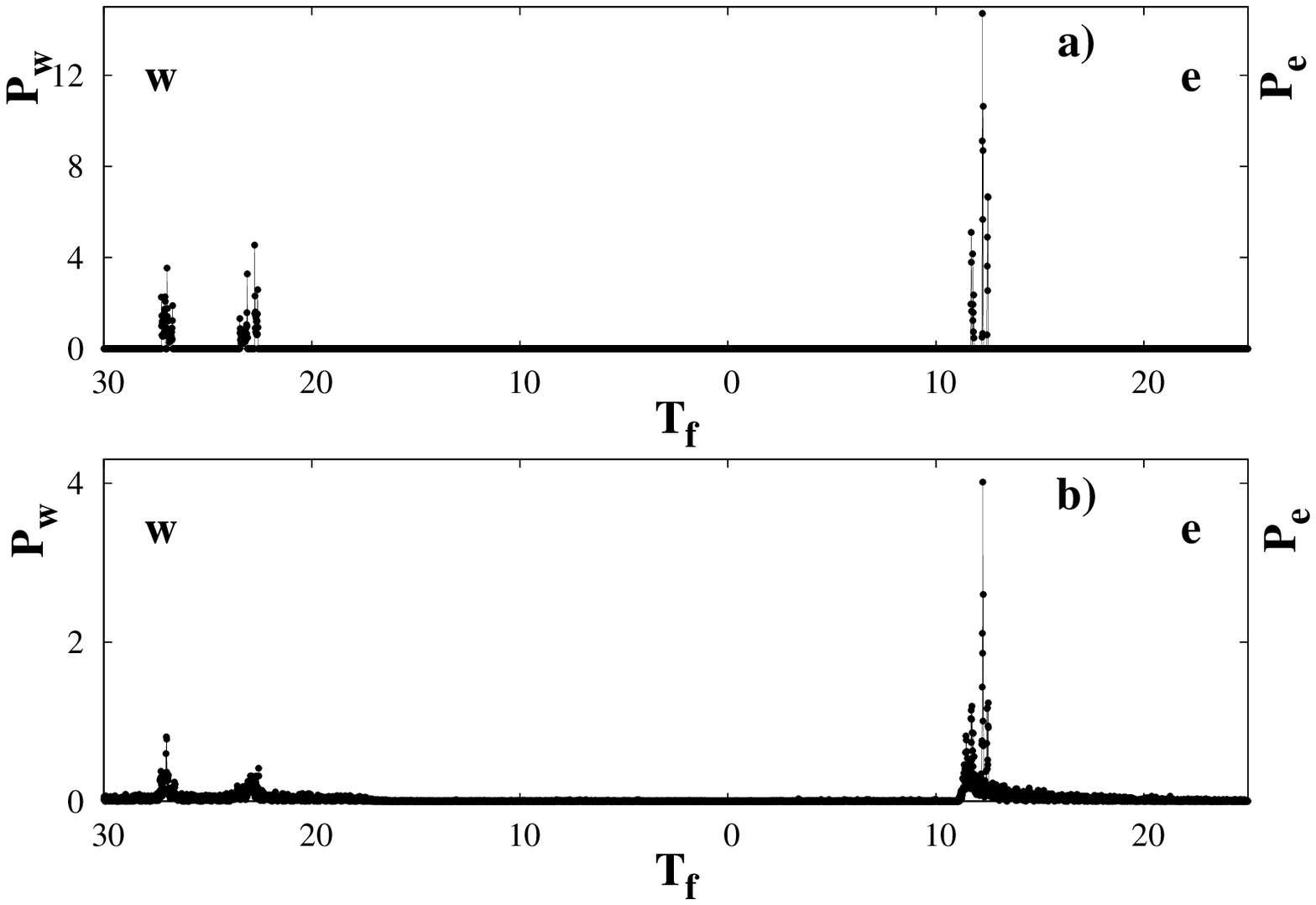}}
\caption{The secondary-resonance islands trap. The PDFs of durations $T_f$
of the eastward (e) and westward (w) flights.  (a) Regular
quasiperiodic trajectory inside the islands with the duration $t=5\cdot 10^5$. 
Both the PDFs are normalized to the number $1.5 \cdot 10^4$ of corresponding flights.  
(b) Chaotic trajectory sticking to the island's boundary from the outside with the duration $t=5\cdot 10^5$. 
$P_{\rm w}(T_f)$ and $P_{\rm e}(T_f)$ are normalized
to the number of westward ($1.1 \cdot 10^4$) and eastward 
($9 \cdot 10^3$) flights, respectively.}
\label{fig7}
\end{figure}

The effect of the RIT of the secondary-resonance islands is illustrated in
Fig.~\ref{fig4}. To find the characteristic times of this RIT we compute two
trajectories: a regular quasiperiodic one with the initial position inside
one of these islands and a chaotic one with the initial position close to
the outer border of the island. The respective PDFs $P(T_f)$, shown in
Figs.~\ref{fig7}~a and b, demonstrate strong correlations between the
corresponding peaks at $T_f\simeq 12, 23$, and $27$.
Computed (but not shown here) PDFs $P(x_f)$ for these trajectories confirm
the effect of the islands RIT on the statistics of short flights.

\section{Saddle traps}

As a result of the periodic perturbation (\ref{5}), the saddle points 
of the unperturbed system  (\ref{4}) at $x_s^{(n)}=2\pi n$, 
$y_s^{(n)}=L\operatorname{Arcosh}
\sqrt{1/LC}+A$  and at $x_s^{(s)}=(2n+1)\pi$, $y_s^{(s)}
=-L\operatorname{Arcosh}\sqrt{1/LC}-A$
($n=0, \pm1, \dots$) become periodic saddle trajectories. These hyperbolic 
trajectories have their own stable and unstable manifolds and play a role of 
specific dynamical traps which we call {\it saddle traps} (ST). In this 
section we demonstrate that the STs influence strongly on chaotic mixing and 
transport of passive particles and contribute, mainly, in the short-time statistics 
of flights. 

\begin{figure}[!htb]
\centerline{\includegraphics[width=0.48\textwidth,clip]{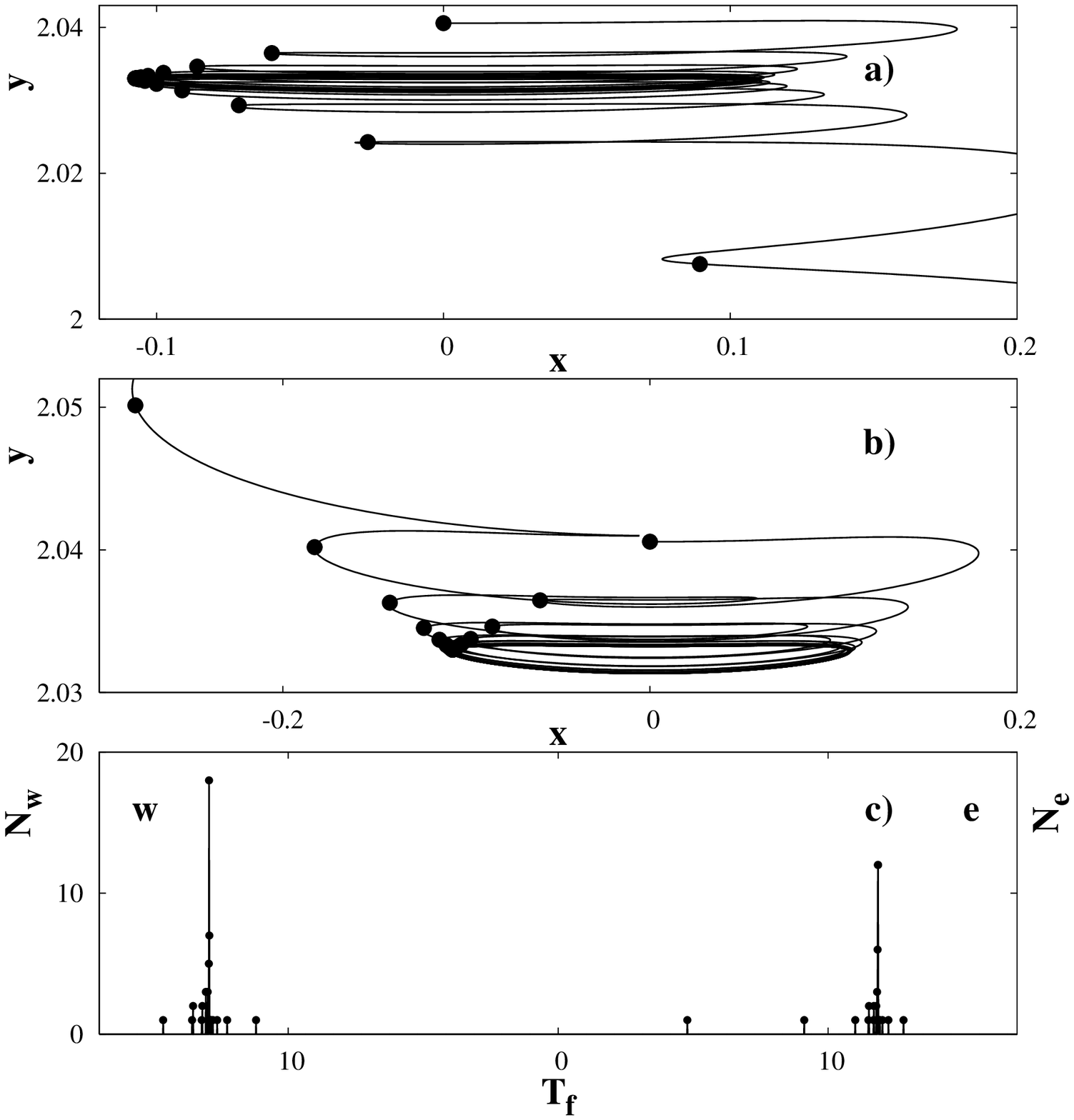}}
\caption{The saddle trap. Fragments of two chaotic trajectories 
sticking to the periodic saddle trajectory one of which escapes to the 
east (a) and another one to the west (b). 
(c) The number of the eastward $(N_{\rm e})$ and westward 
$(N_{\rm w})$ short flights with duration $T_f$ for those two 
trajectories. Statistics with two trajectories with the duration 
$t=10^3$ and the total number of western 
$N_{\rm w}=55$ and eastern $N_{\rm e}=51$ flights.}
\label{fig8}
\end{figure}

Tracers with initial positions close to a stable manifold of a 
saddle trajectory are trapped for a while 
performing a large number of revolutions along it. To illustrate the effect 
of the STs we show in Fig.~\ref{fig8}~a and b fragments of two chaotic 
trajectories sticking to the saddle trajectory and performing about  
20 full revolutions before escaping to the east  (Fig.~\ref{fig8}~a) and to the west 
(Fig.~\ref{fig8}~b). We have managed to detect and locate the corresponding 
periodic unstable saddle trajectory which is situated in 
Figs.~\ref{fig8}~a and b in the domain where a few fragments of the chaotic trajectory imposed 
on each other. Because of the flow asymmetry, the duration 
of eastern flights of a particle along the saddle trajectory 
$T_{\rm e}\simeq 11.9$ is shorter than  the duration 
of western flights $T_{\rm w} \simeq 12.9$. 
The black points are the tracks of the particle's positions 
on the flow plane at the moments of time $t_n=2\pi n/\omega \simeq 24.8\,n$ 
(where $n=1,2,\dots$). They belong to smooth curves which are fragments of the 
stable and unstable manifolds of the saddle trajectory at the chosen initial phase 
$\phi =\pi/2$. 

To estimate the contribution of the STs to the statistics of short flights 
shown in Fig.~\ref{fig5}, we compute and plot in Fig.~\ref{fig8}~c 
the number of the eastward $(N_{\rm e})$ and westward $(N_{\rm w})$
short flights 
with a given duration $T_f$ for those two chaotic trajectories 
sticking to the saddle trajectory arising from the saddle point with 
the position $x_s=0, y_s\simeq 2.02878$. Each full 
rotation of the particles consists of an eastward flight with 
the duration $T_{\rm e} \simeq 11.9$ and an westward flight with 
the duration  $T_{\rm w} \simeq 12.9$. The flights with $T_{\rm e} \simeq 11.9$ 
contribute to the main peak in Fig.~\ref{fig5} and the flights with 
$T_{\rm w} \simeq 12.9$ to ``the wesward'' plateau in that figure.

\begin{figure}[!htb]
\centerline{\includegraphics[width=0.48\textwidth,clip]{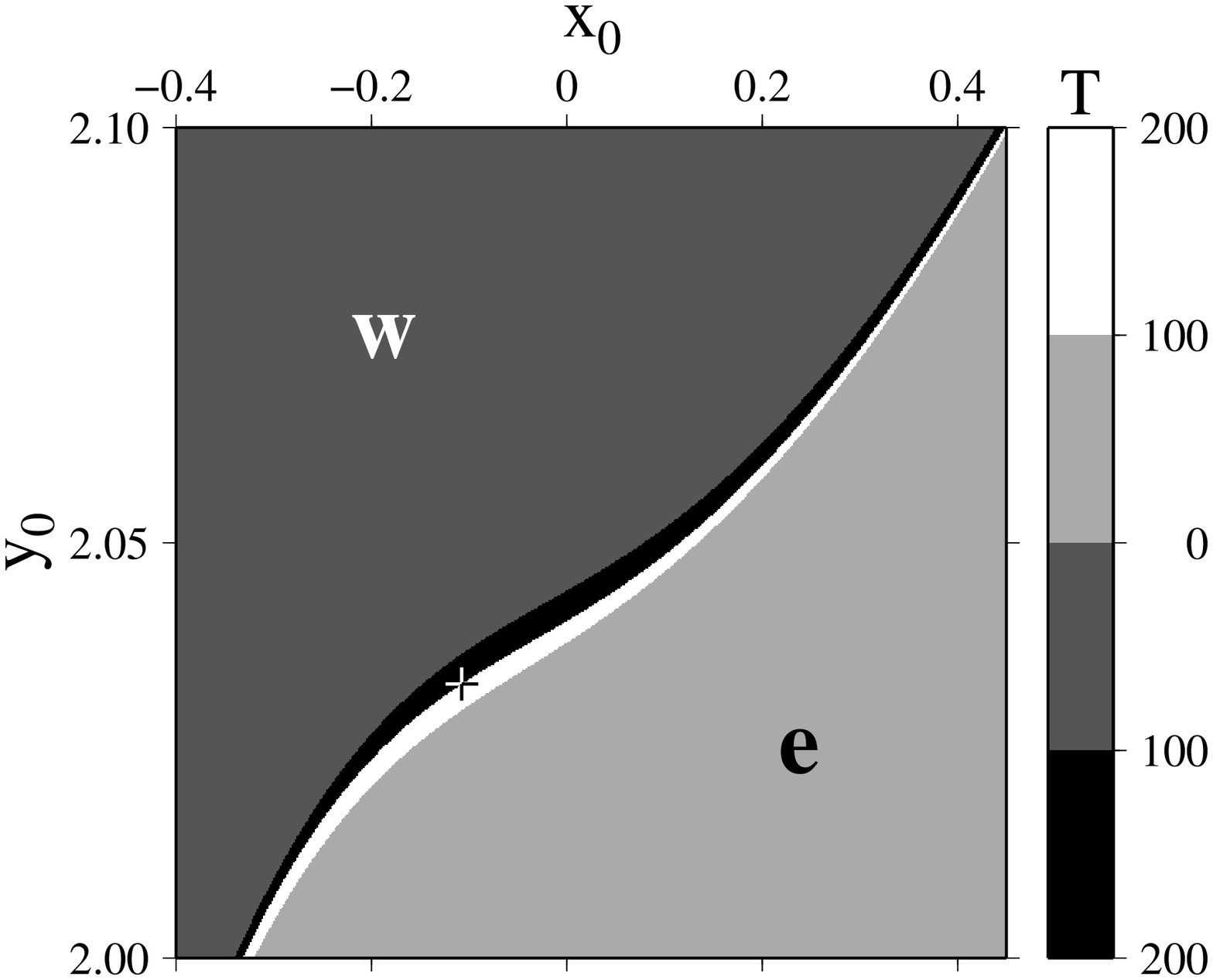}}
\caption{The saddle-trap map. Color modulates the time $T$ which  
$2.5 \cdot 10^5$ particles with 
given initial positions ($x_0,y_0$) need to reach the lines at $x=-1$ or 
$x=1$ escaping to the west (w) and to the east (e), respectively. The  
black and white diagonal curve is an image of a fragment of 
the stable manifold of the 
saddle trajectory. The cross is a position of a particle on that trajectory
at the initial time moment. The integration time is $t=500$.}
\label{fig9}
\end{figure}

The mechanism of operation of the STs can be described as follows. 
Each saddle trajectory $\gamma (t)$ possesses  time-dependent stable $W_s(\gamma (t))$ 
and unstable $W_u(\gamma (t))$ material manifolds composed of a continuous sets of 
points through which pass at time $t$ trajectories of fluid particles 
that are asymptotic to $\gamma (t)$ as $t \to \infty$ and 
$t \to -\infty$, respectively. Under a periodic perturbation, the 
stable and unstable  manifolds  oscillate with the period of the perturbation. 
It was firstly proved by Poincar{\'e} that 
$W_s$ and $W_u$ may intersect each other transversally at an infinite number of 
homoclinic points through which pass doubly asymptotic trajectories. To give 
an image of a fragment of the stable manifold of the periodic saddle
trajectory, we distribute homogeneously $2.5 \cdot 10^5$ 
particles in the  rectangular $[-0.4\le x\le 0.45; 2\le y\le 2.1]$ 
and compute the 
time the particles need to escape the rectangular. The color in Fig.~\ref{fig9} 
modulates the time $T$ when  particles with given initial positions 
$(x_0,y_0)$ reach the western line at $x=-1$ or the eastern line at $x=1$. 
The particles with initial positions marked by the black and white colors 
move close to the stable manifold of the saddle trajectory and spend a maximal 
time near it before escaping. The black and white diagonal curve 
in Fig.~\ref{fig9} 
is an image of a fragment of the corresponding stable manifold.
The particles with initial positions 
to the north from the curve escape to the west along the unstable manifold 
of the saddle trajectory whereas those with initial positions 
to the south from the curve escape to the east along its another unstable manifold. 

\begin{figure}[!htb]
\centerline{\includegraphics[width=0.48\textwidth,clip]{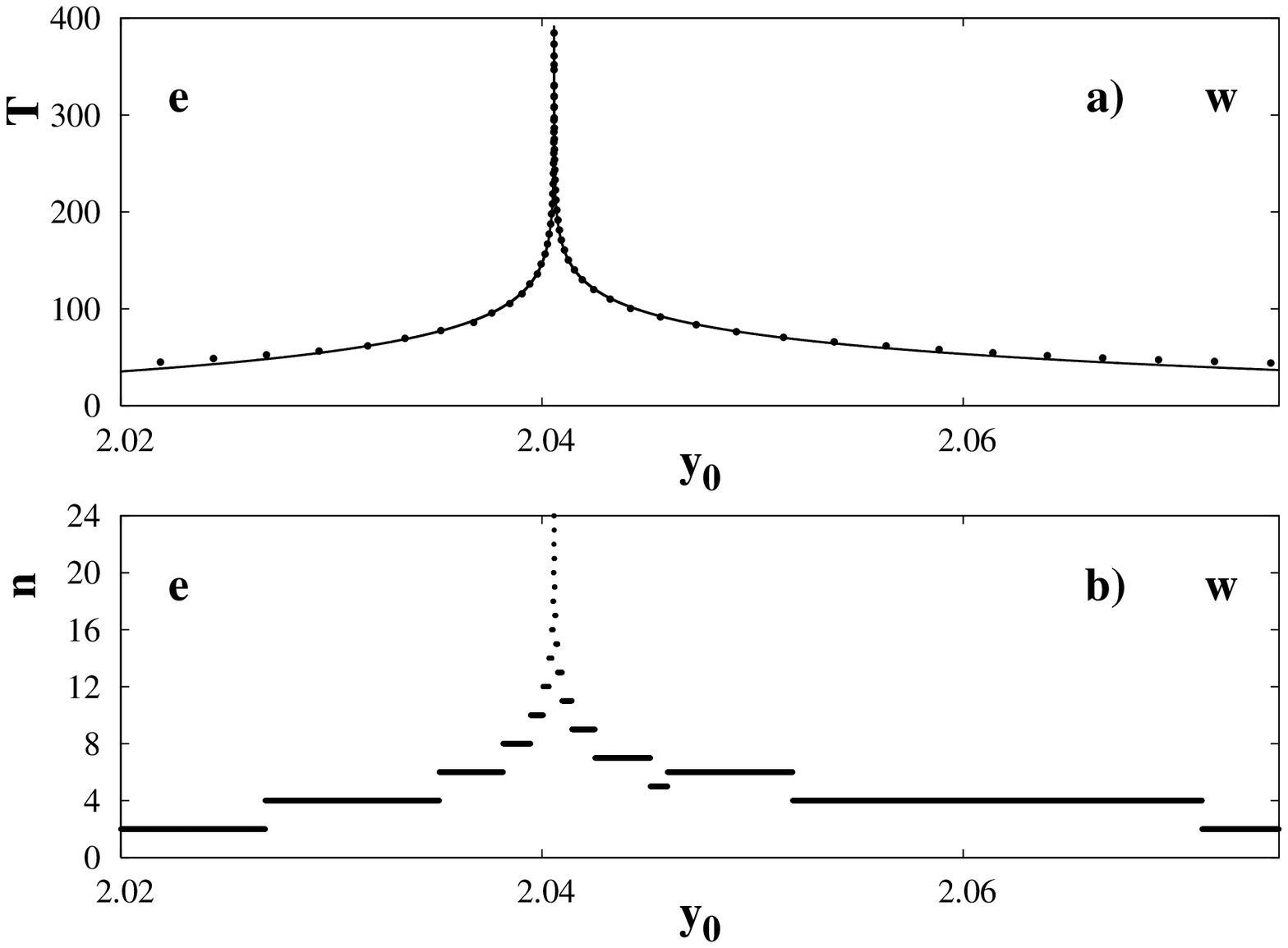}}
\caption{(a) Time $T$ a particle with an initial latitude position $y_0$ 
needs to quit the saddle trap. (b) The number of short flights $n$ such 
a particle performs before quitting the saddle trap. The ranges of $y_0$
from which particles quit the trap moving to the west and east are denoted by 
``w'' and ``e'', respectively.}
\label{fig10}
\end{figure}

We have found that particles quit the ST along the unstable manifolds in accordance 
with specific laws. We distribute a large number of particles along the 
segment with 
$x_0=0$ and $y_0=[2.02;2.06]$, crossing the stable manifold $W_s$, and compute 
the time $T$ particles with given initial latitude positions $y_0$ need to quit 
the ST. More precisely, $T(y_0)$ is a time moment when a particle with 
the initial position $y_0$ reaches the lines with $x=-1$ or $x=1$.  The 
``experimental'' points in Fig.~\ref{fig10}~a fit the law $T_{\rm e}=
(-85.81 \pm 0.04) - 
(31.216 \pm 0.007) \ln(y_{0s}-y_0)$ for the particles which quit the trap moving to 
the east and the law $T_{\rm w}=(-60.61 \pm 0.03) - 
(28.933 \pm 0.006) \ln(y_0-y_{0s})$ for those particles which move to the west when 
quitting the trap, where $y_{0s}=2.0405755472$ is a crossing point of $W_s$ with the segment 
of initial positions. 

The ST attracts particles and force them to rotate in its zone of influence 
performing short flights, the number of which $n$ depends on particle's 
initial positions $y_0$. The $n(y_0)$ is a steplike function (see 
Fig.~\ref{fig10}~b) with the lengths of the steps decreasing in a geometric 
progression in the direction to the singular point, 
$l_j=l_0\,q^{-j}$, where $l_j$ is the length of the $j$-th step and 
$q\simeq 2.27$ for the western exits and $q\simeq 2.20$ for the eastern  ones. 
The seeming deviation from this law in the range $y_0=[2.045;2.046]$ 
(see a small western segment between two larger ones in Fig.~\ref{fig10}~b) 
is explained by crossing the initial 
line $y_0=[2.02;2.06]$ by the curve of zero zonal velocity $u$. To have the 
correct law for the western exits, it is necessary to add the two 
segments of that cut step. The asymmetry of the functions $T(y_0)$ and 
$n(y_0)$ is caused by the asymmetry of the flow. 

\section{Ballistic-islands traps}

Besides the rotational islands with particles moving around the corresponding
elliptic points in the same frame, we have found in Paper~I ballistic islands
situated both in the stochastic layer and in the peripheral currents.
Regular ballistic modes \cite{VRKZ99} correspond to stable quasiperiodic
inter-frame motion of particles. Only the ballistic islands in the stochastic
layer are important for chaotic transport. Mapping positions of the regular ballistic trajectories
at the moments of time $t_n=2\pi n/\omega$ $(n=1,2,\dots)$  onto the first frame, we
obtain chains of ballistic islands both in the northern and southern
stochastic layers, i.~e., between the borders of the northern (southern)
peripheral currents and of the corresponding vortex cores.
A chain with three large ballistic islands is situated in those
stochastic layers. The particles, belonging to these islands, move to the west, and their
mean zonal velocity can be
easily calculated to be $\left< u_f\right>=-2\pi/3T=-\omega/3\simeq -0.0845$.
There are also chains of smaller-size ballistic islands along the very
border with the peripheral currents.

\begin{figure}[!htb]
\centerline{\includegraphics[width=0.48\textwidth,clip]{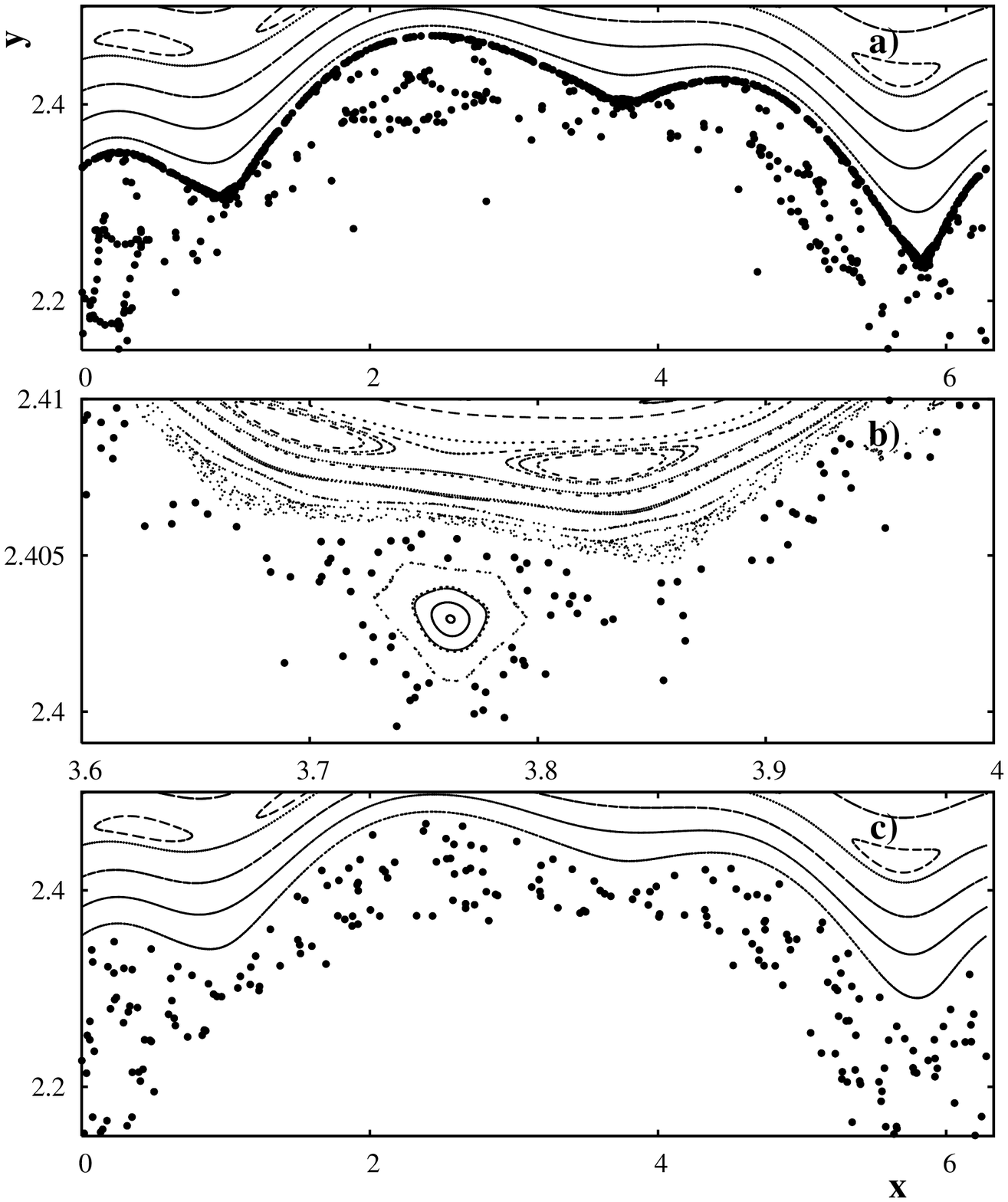}}
\caption{(a) Poincar{\'e} section of the northern stochastic layer 
where stickiness to the very border with the regular westward current
and to three large ballistic islands are shown. Increased density of points along
the border with the peripheral current is caused by the traps of the
border ballistic islands one of which is shown in (b). (c) The trap of the 
large ballistic islands.}
\label{fig11}
\end{figure}

We have demonstrated in Paper~I a stickiness of chaotic trajectories to the
borders of those three large ballistic islands (see Figs.~6 and 7
in Paper~I). The Poincar{\'e} section with fragments of two chaotic
trajectories in the northern stochastic layer
is shown in Fig.~\ref{fig11}~a. One particle performs a long flight sticking
to the very border with the regular westward current, and another one moves to
the west sticking to the very boundaries of three large ballistic 
islands. A magnification of a fragment of the border and tracks of a sticking 
trajectory around a smaller-size ballistic island  are demonstrated in 
Fig.~\ref{fig11}~b. Fig.~\ref{fig11}~c demonstrates the effective size  
of the trap of the large ballistic islands with tracks of a sticking 
trajectory around them.   
\begin{figure}[!htb]
\centerline{\includegraphics[width=0.48\textwidth,clip]{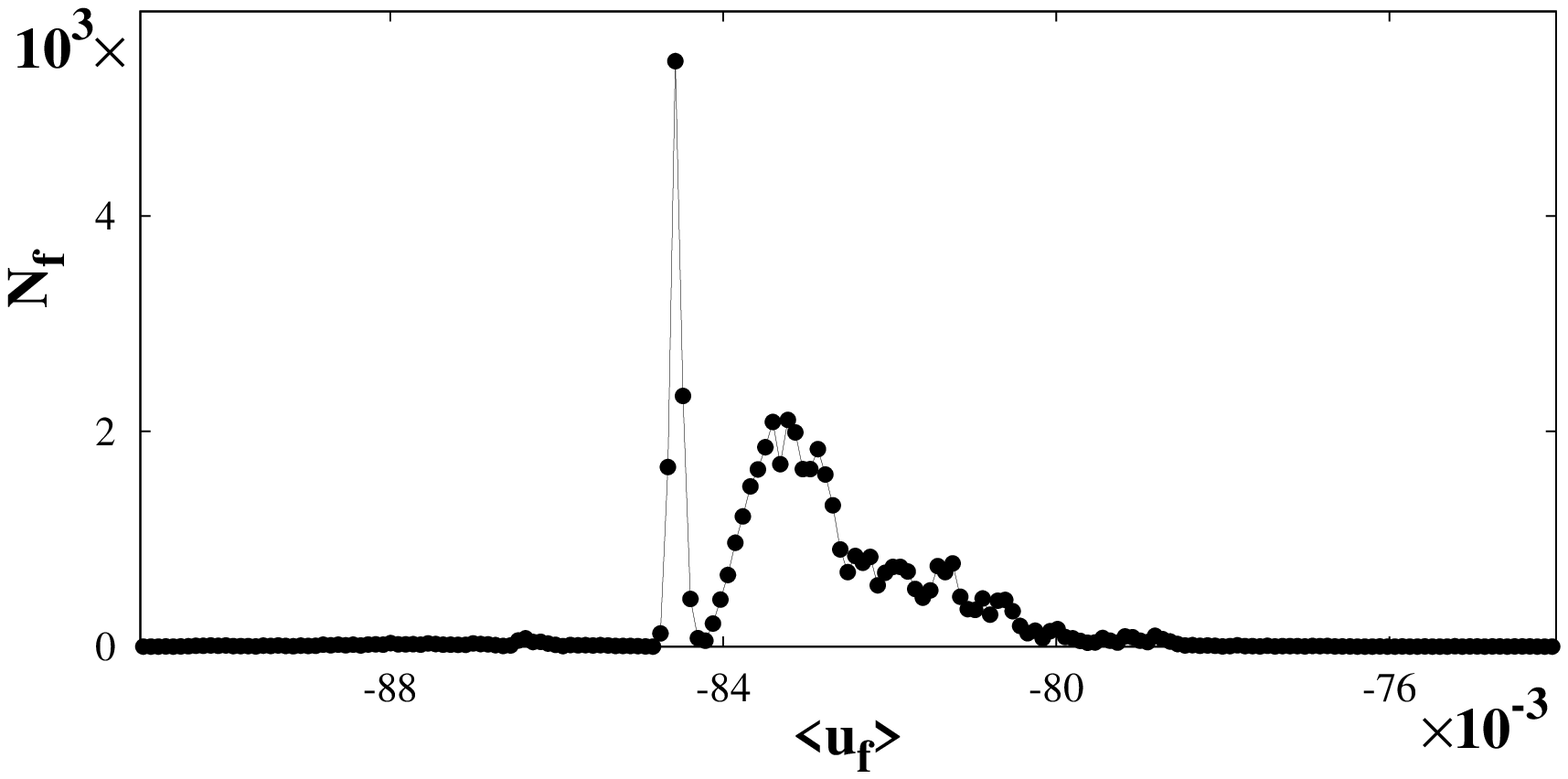}}
\caption{The distribution of a number of long
westward flights with $T_f\ge 10^3$ over their mean  zonal
velocities $\left< u_f\right>$. The sharp peak
corresponds to the trap connected with the very boundaries of the large ballistic
islands, the left wing~--- to a number of traps of families of the border
ballistic islands, and the right wing~--- to the trap situated around
the large ballistic islands. Statistics for five tracers with the total number 
of long westward flights $N_f=5 \cdot 10^4$ and 
the computation time $t=5\cdot 10^8$ for each tracer.}
\label{fig12}
\end{figure}

It is reasonable to suppose that the {\it ballistic-islands traps} (BIT)
contribute, largely, to the statistics of long flights with $|x_f|\gg 2\pi$.
All the ballistic particles, moving both to
the west and to the east, can finish a flight and make a turn only 
in the strip shown in Fig.~\ref{fig1}~c. The loci of the corresponding turning points
have a complicated fractal-like structure. We consider further only long westward flights, taking place
in the northern stochastic layer, because it is much wider than the stochastic layer
between the regular central jet and the southern parts of the vortex cores 
where  eastward flights take place.

To distinguish between contributions of the traps of different
ballistic islands (and, maybe, other zones in the phase space) to the
statistics of long flights, we compute for five long chaotic
trajectories (up to $t=5\cdot 10^8$) the distribution of a number of
westward flights with $T_f\ge 10^3$ over the mean  zonal velocities
$\left< u_f\right>=x_f/T_f$  
of the particles performing such flights. 
The distribution in Fig.~\ref{fig12} has a prominent peak centered at the mean 
zonal velocity $\left< u_f\right>\simeq -0.0845$ which corresponds
to a large number of long flights of those particles (and their
trajectories) which stick to the very boundaries of the large ballistic
islands (see Fig.~\ref{fig11}~a) 
moving  with the mean  velocity $\left< u_f\right>\simeq -0.0845$.
The flat left wing of the distribution $N(\left< u_f\right>)$  corresponds
to the traps of smaller-size ballistic islands nearby the border
with the peripheral current. There are different families of these
islands (see one of them in Fig.~\ref{fig11}~b) with their own values of the mean 
zonal velocity which are in the range
$-0.092\lesssim \left< u_f\right>\lesssim -0.0845$. Stickiness to the
boundaries of the border islands is weaker because they are
smaller than the large islands and their contribution to the
statistics of long flights is comparatively small.

The right wing of the distribution $N(\left< u_f\right>)$
with $-0.084\lesssim \left< u_f\right>\lesssim -0.075$
deserves further investigation. The value
$\left< u_f\right>\simeq -0.075$ is a minimal value of the
zonal velocity for long westward flights possible in the
northern stochastic layer. Increasing the minimal duration of a flight
from $T_f=10^3$ to $T_f=(2\div 5)\cdot 10^3$, we have found splitting of the
broad distribution with
$-0.084\lesssim \left< u_f\right>\lesssim -0.08$ into a number of small distinct
peaks. Comparing trajectories with the values of $\left< u_f\right>$
corresponding to these peaks, we have found that all they move around
the large ballistic islands. The particles with smaller values of $\left< u_f\right>$
used to penetrate further to the south from the islands more frequently
than those with larger values of $\left< u_f\right>$ which prefer to
spend more time in the northern part of the dynamical trap connected with
those islands. Thus, we attribute the right wing
of the distribution $N(\left< u_f\right>)$ to an effect of the trap situated
around the large ballistic islands.

To estimate the contribution of different BITs to the statistics of long westward  
flights in Fig.~\ref{fig2} we have computed the PDFs $P(x_f)$ and $P(T_f)$ 
for particles performing westward flights with $x_f \geq 100$ and 
$T_f \geq 1000$ and with the mean zonal velocity $\left< u_f\right>$ to be 
chosen in three different ranges shown in Fig.~\ref{fig12}:
$-0.092 \lesssim \left< u_f\right>\lesssim -0.085$ 
(particles sticking to the border islands) 
$-0.085 < \left< u_f\right> \lesssim -0.084$ (particles sticking to 
 the very boundary of three large
islands),  and $-0.084 < \left< u_f\right>\lesssim -0.075$ 
(the trap of the three large islands). All the PDFs $P(x_f)$ 
decay exponentially but with different values of the exponents equal to 
$\nu \simeq -0.005$ and $\nu \simeq -0.0018 \div -0.0014$  
for the traps of border and the large ballistic islands, 
respectively. The tail of the PDF $P(x_f)$ for westward flights, 
shown in Fig.~\ref{fig2}, 
decays exponentially with $\nu \simeq -0.0014$. Thus, the contribution of 
the large island's BIT to the statistics of long westward flights is 
dominant.  
As to temporal PDFs $P(T_f)$ for westward long flights, 
they are neither exponential nor power-law 
like with strong oscillations at the very tails. The slope for the border 
BITs is again smaller than for the large ballistic islands trap.

\section{\label{SecConc}Conclusion}

A meandering jet is a fundamental structure in oceanic and atmospheric flows.
We described statistical properties of chaotic
mixing and transport of passive particles in a kinematic model
of a meandering jet flow in terms of dynamical traps
in the phase (physical) space. The boundaries of rotational islands
(including the vortex cores) in circulation zones are dynamical traps (RITs)
contributing, mainly, to the statistics of short flights with $|x_f|<2\pi$.
Characteristic times and spatial scales of the RITs have been shown to
correlate with the PDFs for the lengths $x_f$ and durations
$T_f$ of short flights. The stable manifolds of periodic saddle trajectories 
play a role of saddle traps (STs) with the specific values of the lengths and 
durations of short flights of the particles sticking to the 
saddle trajectories.
The boundaries of ballistic islands
in the stochastic layers (including those situated along the border
with the peripheral current) are dynamical traps (BITs)
contributing, mainly, to the statistics of very long flights with
$|x_f|\gg2\pi$.

Dynamical traps are robust structures in the phase space of  dynamical
systems in the sense that they present at practically all values of the
corresponding control parameters. We never know exact values of the
parameters in real flows, especially, in geophysical ones. We don't know exactly
the structure of the corresponding phase space, however, we know that typical
features, like islands of regular motion, vortices, and jets,
exist in real flows (see their images in some laboratory flows in
Ref. \cite{Ottino}).  In this paper we chose specific values of the
control parameters for which specific PDFs have been computed and explained
by the effect of those  dynamical traps that exist under the chosen parameters.
We have carried out computer experiments with different
values of the control parameters and found that the phase space structure has been
changed, of course, with changing the values of the parameters, but the
corresponding RITs, STs, and BITs with specific temporal and spatial characteristics
have been found to contribute to the corresponding statistics.

After finishing our work, we were acquainted with
Ref. \cite{RW06} where meridional chaotic transport, associated
with a similar kinematic model of a meandering jet, has been
studied by the method of lobe dynamics \cite{W92}.
In difference from our study of zonal chaotic transport, 
a geometric structure of cross
jet transport has been considered in Ref.~\cite{RW06}
where values of the control parameters have been chosen
to be sufficiently large to break up the central jet as a barrier
to transport of particles across the jet. The mechanisms for
particles to cross the jet have been described in terms of lobe
dynamics and the mean  time to cross the jet for particles
entering the jet and the mean  residence time for particles
in the jet have been estimated in Ref.~\cite{RW06}.
We have studied a more realistic situation (at least, for
surface oceanic jet currents) when the jet is an absolute barrier to
cross jet transport and we explained statistical properties
of transport in terms of dynamical traps of saddle trajectories, 
rotational and
ballistic islands. The method of lobe dynamics is hardly applicable for 
study zonal chaotic transport since it is practically impossible to 
trace out lobe evolution for a large number of frames.

\section*{Acknowledgments}
The work was supported by the Russian Foundation
for Basic Research (Grant  no.~06-05-96032), by the Program
``Mathematical Methods in  Nonlinear Dynamics'' of the Russian
Academy of Sciences, and by the Program for Basic Research of
the Far Eastern Division of the Russian Academy of
Sciences.

\end{document}